\documentclass[conference]{IEEEtran}
\IEEEoverridecommandlockouts
\usepackage{cite}
\usepackage{amsmath,amssymb,amsfonts}
\usepackage[ruled,vlined, linesnumbered]{algorithm2e}
\usepackage{graphicx}
\usepackage{textcomp}
\usepackage{xcolor}
\usepackage {bm}
\usepackage{comment}
\usepackage{nicematrix}
\usepackage{tikz}
\usepackage{subcaption}
\def\BibTeX{{\rm B\kern-.05em{\sc i\kern-.025em b}\kern-.08em
T\kern-.1667em\lower.7ex\hbox{E}\kern-.125emX}}
    
\newcommand{\tauazn}{{\tau_{{\rm az}, y}}}
\newcommand{\taueln}{{\tau_{{\rm el}, z}}}
\newcommand{\phiazn}{{\phi_{{\rm az}, y}}}
\newcommand{\phieln}{{\phi_{{\rm el}, z}}}
\def\*#1{\mathbf{#1}}
\newcommand{\figref}[1]{Fig.~\ref{#1}}

\newcommand{\BNg}[1]{\textcolor{red}{BNg: }}

\newcounter{newenumi}
\usepackage[normalem]{ulem} % provides strikeout command \sout

\newcommand{\citep}[1]{\cite{#1}}

\renewcommand{\tilde}{\widetilde}
\renewcommand{\hat}{\widehat}

\def\beq{\begin{equation}}
\def\eeq{\end{equation}}
\def\beqa{\begin{eqnarray}}
\def\eeqa{\end{eqnarray}}
\def\beqan{\begin{eqnarray*}}
\def\eeqan{\end{eqnarray*}}

\DeclareMathOperator*{\argmin}{arg\,min}
\DeclareMathOperator*{\argmax}{arg\,max}

\newcommand\blfootnote[1]{%
  \begingroup
  \renewcommand\thefootnote{}\footnote{#1}%
  \addtocounter{footnote}{-1}%
  \endgroup
}

\begin{document}

\title{3D Beamforming Through Joint Phase-Time Arrays}

\author{\IEEEauthorblockN{Ozlem Yildiz\textsuperscript{1,2}, Ahmad AlAmmouri\textsuperscript{1}, Jianhua Mo\textsuperscript{1}, Younghan Nam\textsuperscript{1}, Elza Erkip\textsuperscript{2}, and Jianzhong (Charlie) Zhang\textsuperscript{1}}
\IEEEauthorblockA{\textsuperscript{1}\textit{Standards and Mobility Innovation Laboratory, Samsung Research America}, Plano, TX 75024, USA}
\textsuperscript{2}\textit{Electrical and Computer Engineering, NYU Tandon School of Engineering, New York University}, Brooklyn, NY, USA\\
\textsuperscript{1}\{ahmad1.a, jianhua.m, younghan.n, jianzhong.z\}@samsung.com, \textsuperscript{2}\{zy2043, elza\}@nyu.edu
}

\maketitle

\begin{abstract}
High-frequency wideband cellular communications over mmWave and sub-THz offer the opportunity for high data rates. However, it also presents high path loss, resulting in limited coverage. High-gain beamforming from the antenna array is essential to mitigate the coverage limitations. The conventional phased antenna arrays (PAA) cause high scheduling latency owing to analog beam constraints, i.e., only one frequency-flat beam is generated. Recently introduced joint phase-time array (JPTA) architecture, which utilizes both true-time-delay (TTD) units and phase shifters (PSs), alleviates analog beam constraints by creating multiple frequency-dependent beams for scheduling multiple users at different directions in a frequency-division manner. One class of previous studies offered solutions with ``rainbow" beams, which tend to allocate a small bandwidth per beam direction. Another class focused on uniform linear array (ULA) antenna architecture, whose frequency-dependent beams were designed along a single axis of either azimuth or elevation direction. This paper presents a novel 3D beamforming design that maximizes beamforming gain toward desired azimuth and elevation directions and across sub-bands partitioned according to scheduled users' bandwidth requirements. We provide analytical solutions and iterative algorithms to design the PSs and TTD units for a desired subband beam pattern. Through simulations of the beamforming gain, we observe that our proposed solutions outperform the state-of-the-art solutions reported elsewhere. 
\blfootnote{This work was done in part while O. Yildiz was an intern at Samsung Research America.}
\end{abstract}

\begin{IEEEkeywords}
True time delay, beamforming, millimeter wave, 3D, joint phase-time array, uniform planar array
\end{IEEEkeywords}

\section{Introduction}

%Action items:
%1. Enhance the problem formulation by relating it to the sum rate.

%2. Add complexity comparison subsection in Simulation. (Ozlem) -Done
%\Ozi{Maybe we should give computing tool specifications? }

%3. Make it clear that we have compared it with prior art.(Ozlem)- Done

%4. Increase to 7 pages.- Done

%5. Enlarge the figures. (Ozlem) - Done

%6. Add the hardware challenges and a holistic view of systems. 

Higher frequency mmWave bands are used in current and next-generation wireless networks due to larger bandwidth  (BW) availability to provide high data rates \cite{andrews2014will}. 
% akyildiz2014terahertz THz band is out of the 6G scope.
However, these frequency bands experience signal degradation due to higher path loss and shadowing effects, which could be resolved using large antenna arrays with directional beamforming \cite{rappaport2019wireless}.

The disadvantage of directional beamforming is that traditional architectures like phased antenna arrays (PAA) cause high scheduling latency owing to analog beam constraints, i.e., only one frequency-flat beam is generated per radio-frequency (RF) chain. Traditional architectures serve different user directions in a time-division-multiplexing (TDM) manner. To serve multiple users in one timeslot, digital beamforming relying on multiple RF chains can be beneficial, but they lead to high power consumption and cost.

We focus on frequency-dependent beamforming architecture, which could be realized by leaky-wave antennas or true-time-delay (TTD) units \cite{forbes2023ttd}. Since leaky-wave antennas are cumbersome and inefficient \cite{lin2022multi}, we consider 
an architecture with TTD units, which is called joint phase time array (JPTA) architecture \cite{ratnam2022joint,alammouri2022extending}. This architecture consists of TTD units and phase shifters (PSs). Using JPTA, beam training can be performed using a single OFDM symbol \cite{boljanovic2021fast}. JPTA can also eliminate the beam squint/split in wideband communications \cite{lin2022multi, dai2022delay}. In addition, JPTA can extend coverage area and increase cell or user throughput by offering more per-user scheduling opportunities thanks to multiple frequency-dependent beams~\cite{alammouri2022extending}. 

Frequency-dependent beamforming could be beneficial during both the initial access and data communication phases. In \cite {boljanovic2021fast} and \cite{wadaskar20213d}, the authors showed that the initial access pilots could be transmitted towards all directions (across azimuth and elevation directions) in one time-domain resource. In \cite{jain2023mmflexible} and  \cite{ratnam2022joint}, the data communication phase is improved by steering the frequency-dependent 2D beams to multiple users' angle of arrivals (AoAs). This way, multiple users in different directions are scheduled in a frequency-division-multiplexing (FDM) manner, even when the antenna panel is associated with only one RF chain.

In this work, we propose methods to derive a 3D frequency-dependent beam by optimizing the phase and delay parameters in the JPTA architecture. The formed beam, which is used for data communications, considers the UE locations and their BW requirements, i.e., UEs with higher BW requirements observe high beam gain over a larger BW than UEs with smaller desired BW. A summary of our contributions is as follows:
\begin{itemize}
    \item We analytically derive a closed-form solution for 3D beamforming design by defining the phase and delay values in two different ways: joint and separated according to their interdependence, see Sec.~\ref{sec:joint} and \ref{sec:sep_analytical}.
    \item We improve the performance beyond analytical solutions by applying greedy and gradient descent algorithms which take the analytically derived values as initial values. See Sec.~\ref{sec:greedy} and \ref{sec:gradient}.  
    \item We provide 3D beamforming gain simulations to compare the performance of our methods with the 3D extension of the state-of-the-art 2D beamforming codebook design in \cite{ratnam2022joint}. We demonstrate that our iterative algorithms outperform the state-of-the-art by providing more resilient results over different scenarios, see Sec.~\ref{sec:sim}. 
\end{itemize}

\section{System Model}

\begin{figure}[t]
\centering
\begin{subfigure}{\linewidth}
    \centering
\includegraphics[width=0.99\linewidth]{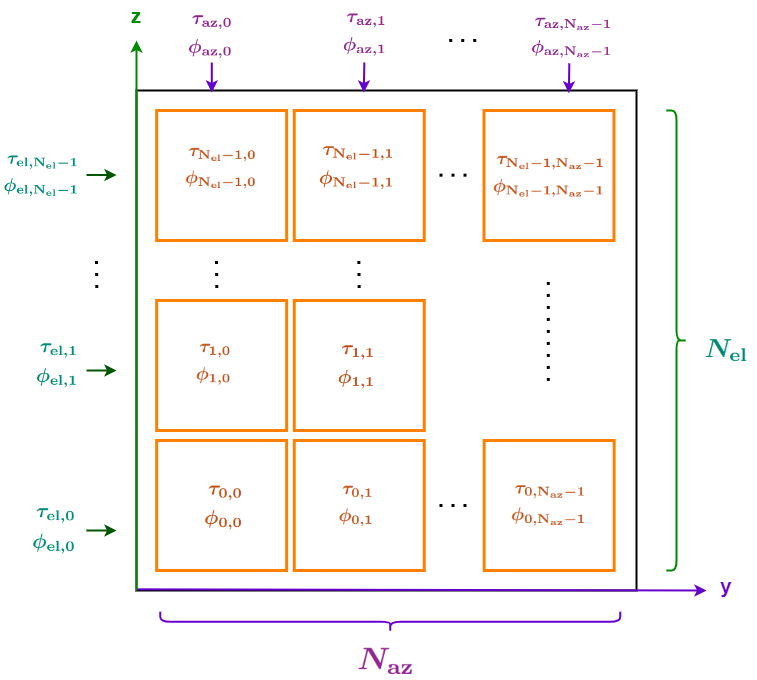}
\end{subfigure}
    \caption{Demonstration of a joint phase-time array and the joint and separated beam design approaches.}
    \label{fig:antenna_design}
\end{figure}

\begin{table}[]
\centering
\begin{tabular}{ll}
\hline
\textbf{Parameters} & \textbf{Values} \\ \hline
Center frequency, $f_c$              & 28 GHz              \\
Bandwidth, {$\rm BW$}                 & 95 MHz             \\
Subcarrier spacing, {$\Delta f$}                    & 120 kHz             \\
Number of subcarriers, $M+1$                & 793           \\
Number of antennas in $y$ direction, $N_{\rm az}$           & 16                  \\
Number of antennas in $z$ direction, $N_{\rm el}$           & 24                  \\
Delay quantization step, $\tau_p$           & 2.5 ns              \\
Delay range, $\tau_{\max}$           & 200 ns              \\
Phase shifter precision, $\beta$            & 6 bits              \\ \hline
\end{tabular}
\caption{Parameter definition and default values.}
\label{tab:tab1}
\end{table}

We consider 3D beam design for uplink communication with a system bandwidth ${\rm BW}$ around a center frequency $f_c$. A single base station (BS) and $N_{ u}$ users are in the environment. The BS is equipped with a single RF chain and a rectangular antenna array, comprising $N_{\rm az}$ antenna elements in the azimuth direction and $N_{\rm el}$ antenna elements in the elevation direction, as shown in \figref{fig:antenna_design}. The antenna spacing is half-wavelength at the center frequency $f_c$. The antenna array steering value for $(y,z)$-th antenna element is as follows
    \begin{equation}
    \begin{aligned}
        \left[\bm a_{R} (\theta_{\rm az}, \theta_{\rm el}, f_m)\right]_{y,z} &= e^{j \pi\frac{f_m}{f_c} (y\sin\theta_{\rm az}\sin\theta_{\rm el} +  z\cos\theta_{\rm el})}, \\
    \end{aligned}
    \label{eq:antenna}
    \end{equation}
where $f_m$ is the subcarrier frequency, $\theta_{\rm az}$ and $\theta_{\rm el}$ denote the AoA in the azimuth and elevation plane. Here, $[\cdot]_{y,z}$ represents the $(y,z)$-th element of the matrix.

\subsection{Joint Phase Time Array Architecture}
The JPTA architecture is shown in \figref{fig:antenna_design}, where every antenna element in the rectangular array has a separate TTD and PS, enabling control over the angle and frequency domains.

\begin{comment}

\begin{figure*}[t]
\centering
\begin{subfigure}{0.31\linewidth}
    \centering
    \includegraphics[width=\textwidth]{Figures/antenna_design.png}
    \caption{}
    \label{fig:antenna_design}
\end{subfigure}
\begin{subfigure}{0.31\linewidth}
    \centering
    \includegraphics[width=\textwidth]{Figures/Upd_Figures/analytical_alpha.eps}
    \caption{}
    \label{fig:analytical_alpha}
\end{subfigure}
\begin{subfigure}{0.31\linewidth}
    \centering
    \includegraphics[width=\textwidth]{Figures/Upd_Figures/uneq_analytical.eps}
    \caption{}
    \label{fig:uneq_analytical.eps}
\end{subfigure}
\caption{}
\label{fig:set_analytical}
\end{figure*} 
\end{comment}

The beamforming combiner of JPTA architecture is
%     \begin{equation}
%         \bm\omega (f_m,\bm\phi,\bm\tau) = \frac{1}{\sqrt{N_{\rm az} N_{\rm el}}} \bm\omega_{\text{phase}}(\bm\phi)  \odot \bm\omega_{\text{delay}} ( f_m,\bm\tau),
%     \end{equation}
% where $\odot$ is an element-wise multiplication and $\bm\tau$ $(\bm\phi)$ corresponds to a matrix containing each antenna element's delay (phase) value. The PS weights are $\left[\bm\omega_{\text{phase}}\right]_{y, z} = e^{j \phi_{y,z}}$ for a given phase $\phi_{y,z}$ and TTD combiner weights are
% $\left[\bm \omega_{\text{delay}} (f_m) \right]_{y,z}  = e^{-j2\pi f_m \tau_{y,z}}$ 
% for a given delay~$\tau_{y,z}$.
    \begin{equation}
        \left[\bm\omega (f_m,\bm\phi,\bm\tau)\right]_{y,z} = \frac{1}{\sqrt{N_{\rm az} N_{\rm el}}} e^{j (\phi_{y,z} + 2 \pi f_m \tau_{y, z})},
    \end{equation}
where $\bm\tau$ $(\bm\phi)$ corresponds to a matrix containing each antenna element's delay (phase) value. The PS weights are $e^{j \phi_{y,z}}$ for a given phase $\phi_{y,z}$ and TTD combiner weights are
$e^{j2\pi f_m \tau_{y,z}}$ 
for a given delay~$\tau_{y,z}$.

The beamforming gain is therefore given as
\begin{equation}
    \begin{aligned}
        &G(\theta_{\rm az}, \theta_{\rm el}, f_m,\bm\phi,\bm\tau) \\
        =& \lVert\bm   a_{R}^H (\theta_{\rm az}, \theta_{\rm el}, f_m ) 
        \bm\omega ( f_m,\bm\phi,\bm\tau)
\rVert^2 \\
        =&  \frac{1}{N_{\rm az}  N_{\rm el}} \left\lVert \sum_{y=0}^{N_{\rm az}-1} \sum_{z=0}^{N_{\rm el}-1} e^{j h( f_m,\phi_{y,z},\tau_{y,z}) -j \Omega(y,z,f_m,\theta_{\rm az}, \theta_{\rm el}) } \right \rVert^2,
    \end{aligned}
    \label{eq:gain}
\end{equation}
%where the delay is $\tau_{ y,z}$, and phase is $\phi_{ y,z}$. The user's AoA is $(\theta_{\rm az}, \theta_{\rm el})$ and the assigned subcarrier frequency is $f_m$. 
with the definition  of $h(f_m,\phi_{y,z},\tau_{y,z}) \triangleq \phi _{ y,z} + 2\pi f_m \tau_{y,z}$ and 
 $\Omega(y,z,f_m,\theta_{\rm az}, \theta_{\rm el}) \triangleq  \pi \frac{f_m}{f_c} (y \sin \theta_{\rm az} \sin \theta_{\rm el} +  z\cos \theta_{\rm el})$ for simplification.

\subsection{Problem Formulation}
Recall that our goal is to maximize the 3D frequency-dependent beamforming gain of $N_u$ users for a given BW allocation of each user by configuring the PSs and the TTD units. The desired BW ratio of the $i$-th user is denoted as $\alpha_i$, which corresponds to $\lfloor \alpha_i \times \frac{\rm BW}{\Delta f} \rfloor$ subcarriers where ${\Delta f}$ is the subcarrier spacing. Denote the assigned subcarrier set to $i^{\rm th}$ user as $\mathcal{F}_{m,i}$, and
the AoAs as $(\theta_{{\rm az},i}, \theta_{{\rm el}, i})$.
%To fully utilize the bandwidth, the equality of $\sum_{i=1}^{N_u}\lfloor \alpha_i \times {\rm BW} \rfloor = {\rm BW}$ is a necessary condition. 
%It's essential that $\alpha_i$s are chosen accordingly to fully utilize the bandwidth, e.g. $\sum_{i=1}^{N_u} \alpha_i \times {\rm BW} = {\rm BW}$. 
%
The average beamforming gain of $i^{\rm th}$ user over the assigned subcarrier set $\mathcal{F}_{m,i}$ at the desired direction $(\theta_{{\rm az},i}, \theta_{{\rm el}, i})$ is thus
\begin{equation}
    G_{{\rm mean}, i} \triangleq \frac{1}{|\mathcal{F}_{m,i}|}\sum_{m\in \mathcal{F}_{m,i}} G(\theta_{{\rm az},i}, \theta_{{\rm el}, i}, f_m). 
    \label{gmean}
\end{equation}

To assess the performance of the beam, we use the \textit{log-mean} of the average user gain as a metric, 
\begin{equation}
    G_{l}  \triangleq
        \sum_{i=1}^{N_u} 10\log_{10}G_{{\rm mean}, i}.
    \label{eq:gl}
\end{equation} 
The utilization of logarithms in this context further promotes fairness among various users \cite{kelly1998rate}. In addition, the formulation approximates maximizing the total sum rate across the $N_u$ users at high SNR since the data rate increases logarithmically with SNR.
Finally, the problem formulation is to maximize the above metric over the delay and phase shifter values:
%$\bm\alpha = [ \alpha_1, \cdots, \alpha_{N_u}]$, which leads to $\alpha_i \times BW$ bandwidth assignment to $i$-th user when the total bandwidth is $BW$. Note that $\sum_{i=1}^{N_u} \alpha_i = 1$, therefore, we use the whole bandwidth. Therefore, the objective can be stated as
\begin{equation}
        \left\{\bm\tau_{\rm}^*, \bm \phi_{\rm}^*\right\} = \argmax_{\bm\tau_{\rm}, \bm \phi_{\rm}}  G_l. 
        \label{eq:pf}
\end{equation} 
%Our objective is to maximize each user's beamforming gain while maintaining a consistent configuration for both $\bm\phi$ and $\bm\tau$ as these values are allocated to each antenna element's PS and TTD unit. 

%where $\bm\theta_{\rm az}, \bm\theta_{\rm el}  \in \mathcal{R}^{N_u}$ are the angle of arrivals of the users and $\mathcal{F}_m$ denotes the subcarriers assigned for the users. It is important to note that the azimuth and elevation gains are not dependent on each other. Therefore, their solutions are also independent. 

\section{3D Beamforming Design}
\label{sec:method}
We provide several methods to design the 3D beam by solving the problem defined in \eqref{eq:pf}. First, we outline an analytical method to jointly determine the phase and delay values in azimuth and elevation directions, referred to as \textit{joint}, in Sec.~\ref{sec:joint}. Next, we define a similar analytical solution when the phase and delay values in the azimuth and elevation directions are designed separately, referred to as \textit{separated}, in Sec.~\ref{sec:sep_analytical}. 
%when the phase and delay values are written as a summation of parameters defined in azimuth and elevation directions, which is described as separated analytical derivation.
Consequently, we propose two iterative optimization techniques to enhance the analytical solutions: denoted as the \textit{greedy} algorithm in Sec.~\ref{sec:greedy} and the \textit{gradient descent} algorithm, in Sec.~\ref{sec:gradient}.

\subsection{Joint Analytical Solution}
\label{sec:joint}
In this section, we devise a solution for the problem outlined in \eqref{eq:pf} by establishing a set of linear equations corresponding to each antenna element, similar to \cite{jain2023mmflexible}, given that the phase and delay of the antenna elements are set independently. 

Looking back at \eqref{eq:gain}, we can see that the maximum beamforming gain for a given direction $(\theta_{\rm az}, \theta_{\rm el})$ and subcarrier $(f_m)$ is achieved when  $h(\cdot) = \Omega(\cdot) $ for all $y$ and $z$ in $[0, N_{\rm az}-1]$ and $[0, N_{\rm el}-1]$, respectively. Hence, if there is a solution satisfying this equality for each UE over its assigned subcarriers, the maximum beamforming gain is achieved for all UEs. 

For analytical convenience, we assume that there are $M+1$ subcarriers and we assign the subcarriers in range $[f_{\alpha_{i-1}(M+1)},f_{\alpha_{i}(M+1)} )$ to $i$-th user. 
Therefore, we formulate a system of linear equations that encompasses an equation for each subcarrier given by $\bm A\bm x_{ y,z} = \bm b_{ y ,z}$, with $\bm x_{ y,z} \triangleq [\phi_{ y,z}, 2\pi \Delta f \tau_{ y,z}]^T$, and 
% \begin{equation}
%     \*A \triangleq
%     \begin{bmatrix}
%         1 & -M/2 \\
%         1 & -M/2+1 \\
%         \vdots & \vdots \\
%         1 & M/2
%     \end{bmatrix}.
% \end{equation}

\begin{equation}
    \bm A \triangleq
    \begin{bmatrix}
        1 & 1 & \cdots & 1 \\
        -M/2 & -M/2+1 & \cdots & M/2 \\
    \end{bmatrix}^T.
\end{equation}

Under the assumption of $f_m/f_c \approx 1$, $\bm b_{y,z}$ is defined as,
\begin{equation}
\bm b_{ y,z} \triangleq [\bm b^T_{y, z, 1}, \bm b^T_{y, z, 2}, \cdots, \bm b^T_{y, z, N_u}]^T,
\end{equation}
where
\begin{equation}
\bm b_{y, z, i} \triangleq (2\pi k(  y,z,i) + \nu( y,z,i)) \begin{bNiceMatrix}%
[margin,
nullify-dots,
xdots/line-style=loosely dotted,
]
1, 1, & \Cdots &, 1  
\CodeAfter
  \UnderBrace[yshift=1mm,shorten]{last-1}{last-3}{\scriptstyle \alpha_i (M+1)}
  %\UnderBrace[yshift=1mm,shorten]{last-4}{last-last}{\scriptstyle M/2 +1}
\end{bNiceMatrix}^T,
\end{equation}
\vspace{0.4cm}

%\JMo{Ozlem, could you confirm if my revision of the above two equations is right.}
%
\noindent where $\nu( y,z,i) \triangleq y \pi  \sin \theta_{ {\rm az}, i} \sin \theta_{{\rm el},i} + z \pi  \cos \theta_{{\rm el},i}$ is the antenna array steering value for the $i$-th user and 
\begin{equation}
    \begin{split}
    & k( y,z, i)  \\
    &\triangleq \begin{cases}
        0 & i=1 \\
        k( y,z, i-1) + { round} \left(\frac{\nu( y,z, i-1) - \nu( y,z, i)}{2\pi}  \right) & {\rm else, }
    \end{cases}
    \end{split}
\end{equation}
is the offset to ensure the gap between the array steering values of different users is less than $2\pi$. 

The defined linear system of equations leads to an over-determined linear system of equations, so finding a unique solution to satisfy every equation is impossible. Thus, we introduce the error term, denoted as $\bm e_{y,z} \triangleq \bm A\bm x_{y,z} -\bm b_{y,z}$. To maximize the beamforming gain, we need to minimize the error $\bm e_{y,z}$ for all combinations of $y$ and $z$. To fulfill this, we explore two approaches by employing different norm values for error quantification, such as the square and infinity norms. 

\subsubsection{Least Squares Solution}
We redefine the problem formulation as  $\min_{\bm\tau,  \bm\phi} \lVert\bm e_i (\bm\theta_{\rm az}, \bm\theta_{\rm el}, \mathcal{F}_m) \rVert_2$
%\begin{equation}
 %       \min_{\bm\tau,  \bm\phi} \lVert\*e_i (\bm\theta_{\rm az}, \bm\theta_{\rm el}, \mathcal{F}_m) \rVert^2,
 %       \label{eq:ls}
%\end{equation}
and by definition, the closed form solution is $\hat{\bm x_{y,z}} = (\bm A^T\bm A)^{-1}\bm A^T\bm b_{y,z}$, which leads to the phase values of
    \begin{equation}
        \phi_{y,z} = \sum_{i=1}^{N_u} \alpha_{i} (\nu(y,z,i)+ 2\pi k(y,z,i)),
        \label{eq:phi}
    \end{equation}
and delay values of
    \begin{equation}
        \tau_{y,z} = \sum_{i=1}^{N_u} (\nu(y,z,i) + 2\pi k(y,z,i) )M_{\rm sum}(i, \bm\alpha),
        \label{eq:tau}
    \end{equation}
where, 
\begin{equation}
    M_{\rm sum} (i, \bm\alpha) \triangleq   \sum_{m=0}^{\sum_{l=1}^{i} \alpha_l M -M/2} m -  \sum_{m=0}^{\sum_{l=1}^{i-1} \alpha_l M -M/2} m.
\end{equation}

\subsubsection{Infinity Norm Solution}
In this part, we define the problem as $\min_{\bm\tau,  \bm\phi} \lVert \bm e_i (\bm\theta_{\rm az}, \bm\theta_{\rm el}, \mathcal{F}_m) \rVert_\infty$. However, this approach does not lead to a closed-form solution, and the optimization can be solved by using a linear programming toolbox, for example, \textit{CVX}.

\subsection{Separated Analytical Solution}
\label{sec:sep_analytical}
We can rewrite the antenna array steering vector, which is given in Eq.~\eqref{eq:antenna}, by the Kronecker product, $\otimes$ of azimuth and elevation antenna array steering vectors as in 
\begin{equation}
    \begin{aligned}
         \bm a_{\rm R} (\theta_{\rm az}, \theta_{\rm el}, f_m) &= \bm a_{\rm az} (\theta_{\rm az}, \theta_{\rm el}, f_m) \otimes \bm a_{\rm el} (\theta_{\rm el}, f_m),\\
        \left[\bm a_{\rm az} (\theta_{\rm az}, \theta_{\rm el}, f_m) \right]_{y}&= e^{j \pi\frac{f_m}{f_c} y\sin\theta_{\rm az}\sin\theta_{\rm el}}, \\
        \left[\bm a_{\rm el} ( \theta_{\rm el}, f_m) \right]_{z} &= e^{j \pi\frac{f_m}{f_c}  z\cos\theta_{\rm el}}.
    \end{aligned}
\end{equation}

This encourages us to rewrite the delay (phase) values in terms of the delays (phases) in azimuth and elevation direction, similar to \cite{wadaskar20213d}. Therefore, the delay at $(y,z)$-th antenna can be expressed as $\tau_{y,z} = \tau_{{\rm el},y} + \tau_{{\rm az}, z}$, in which $\tauazn$ and $\taueln$ are the delays at each antenna element in $y$ and $z$ direction. Similarly, $\phi_{y,z} = \phi_{{\rm el},y} + \phi_{{\rm az}, z}$, in which $\phiazn$ and $\phieln$ are the programmable antenna phases in $y$ and $z$ direction. For a more detailed illustration, please refer to \figref{fig:antenna_design}. 

The definitions above result in revising the beamforming combiner as follows
\begin{equation}
    \begin{aligned}
        \bm\omega ( f_m) 
        %&=  e^{j (\phiazn + 2\pi f_m \tauazn ) + j (\phieln + 2\pi f_m  \taueln ) } \\
        &=  \bm\omega_{\rm az} (f_m) \otimes \bm\omega_{\rm el} ( f_m), \\
\left[\bm\omega_{\rm az} (f_m) \right]_y &= \frac{1}{\sqrt{N_{\rm az}}} e^{j (\phiazn + 2\pi f_m \tauazn ) }, \\
 \left[\bm\omega_{\rm el} (f_m) \right]_z &= \frac{1}{\sqrt{N_{\rm el}}}e^{j (\phieln + 2\pi f_m \taueln ) }. \\       
    \end{aligned}
\end{equation}

By using the above descriptions, the beamforming gain can be expressed as $G(\theta_{\rm az}, \theta_{\rm el}, f_m) = G_{\rm az} (\theta_{\rm az}, \theta_{\rm el}, f_m) G_{\rm el} (\theta_{\rm el}, f_m)$, where
\begin{equation}
    \begin{aligned}
        G_{\rm az} (\theta_{\rm az}, \theta_{\rm el}, f_m) 
        &= \frac{1}{N_{\rm az}} \left\lVert \sum_{y=0}^{N_{\rm az}-1} e^{j h_{{\rm az}}(y)  -j\Omega_{{\rm az}}(y)  } \right \rVert^2,
    \end{aligned}
    \label{eq:azgain}
    \end{equation}
with the definitions of $h_{{\rm az}}(y) \triangleq \phiazn + 2\pi f_m \tauazn$ and $\Omega_{{\rm az}}(y) \triangleq \pi \frac{f_m}{f_c} y \sin \theta_{\rm az} \sin \theta_{\rm el} $, and 
\begin{equation}
    \begin{aligned}
        G_{\rm el} (\theta_{\rm el}, f_m) 
        &= \frac{1}{N_{\rm el}}\left\lVert\sum_{z=0}^{N_{\rm el}-1} e^{j h_{{\rm el}}(z) -j \Omega_{{\rm el}}(z) }\right \rVert^2,
    \end{aligned}
    \label{eq:elgain}
    \end{equation}
with the definitions of $h_{{\rm el}}(z) \triangleq \phieln + 2\pi f_m \taueln$ and $\Omega_{{\rm el}}(z) \triangleq \pi \frac{f_m}{f_c} z \cos \theta_{\rm el}$.

The linear system of equations is formed similarly to Sec.~\ref{sec:joint}. However, we individually engage in the solution of $y$-th antenna with the specific objective of optimizing $G_{\rm az}$, and $z$-th antenna with the intent of maximizing $G_{\rm el}$ since these two gain values are inherently independent of one another, as per their definitions. Consequently, this alternate approach requires $N_{\rm az} + N_{\rm el}$ equations. At the same time, joint configuration necessitates $N_{\rm az} \times N_{\rm el}$ due to individually interacting with each antenna element. 
%Therefore, the separated solution brings a significant saving in the computational complexity at the cost of the achieved beamforming gain, which will be shown in the simulation results.

%engaging in the resolution of linear systems of equations for each antenna element.

\begin{comment}

\begin{figure*}[t]
    \begin{subfigure}{0.31\linewidth}
    \centering
    \includegraphics[width=\textwidth]{Figures/Upd_Figures/comp_final.eps}
    \caption{Various ${\rm BW}$ allocations}
    \label{fig:comp_final.eps}
\end{subfigure}
\caption{Performance comparison of methods in joint and separated optimization under different ${\rm BW}$ allocations }
\label{fig:set_methods}
\end{figure*}
\end{comment}

\subsection{Greedy Algorithm}
\label{sec:greedy}
The objective of the greedy algorithm is to explore the solution space to identify any potential enhancements in the beamforming gain, after analytical derivations. Given that we have established two distinct solution sets for $\bm \tau$ and $\bm \phi$ through the joint and separated optimization techniques, we will subsequently pursue two different greedy optimization approaches: the joint and separated. For both cases, parameters are initialized by using the least squares method because this method results in a closed-form solution. 

\subsubsection{Joint}
Our proposed Algorithm~\ref{alg:1} takes system parameters as input and provides the phase and delay configurations, $\bm \phi$ and $\bm \tau$ as output, respectively. The parameters are initially set based on the equations provided in \eqref{eq:phi} and \eqref{eq:tau}. Subsequently, quantization is performed by defining a specific precision for delay denoted as $\tau_p$ and specifying the bit resolution for phase, indicated as $\beta$. 

Within the two-dimensional antenna array, our algorithm explores the delay grid for a specified antenna element to improve $G_l$ while holding other antenna elements' constant delay and phase values. Subsequently, a similar process is applied to the phase values. These updates continue until a predefined convergence criterion is satisfied. The criterion is the absolute difference in $G_l$ between consecutive runs being smaller than $\zeta$ of the last iteration. We explored the various $\zeta$ values to find the optimal one. Once the convergence criteria are met, the algorithm returns the updated $\bm \phi$ and $\bm \tau$.

\begin{algorithm}
\DontPrintSemicolon
\SetNoFillComment
\caption{Joint Greedy Optimization of  $\bm\phi$, $\bm\tau$}\label{alg:1}
\textbf{Input:} $N_{\rm az}$, $N_{\rm el}$, $M$, $\bm\theta_{\rm az}$, $\bm \theta_{\rm el}$, $\bm\alpha$\;
\textbf{Output:}  $\tilde{\bm\tau}$, $\tilde{\bm\phi}$ \;
Initialize $\bm\phi$ and $\bm\tau$   by using Eqs. \eqref{eq:phi} and \eqref{eq:tau} \;

%\tcc{$\tau_p$ is the delay precision}
Define the candidate delays as $\tau_{\rm grid} = 0: \tau_p : \tau_{\max}$\;

%\tcc{ Phase is defined by $\beta$ bits }
Define the possible phase values as $\phi_{\rm grid} = 0: \frac{2\pi}{2^\beta} : 2\pi $\;

 %Quantize  $\tauazn$, $\taueln$ to the closest value in $\tau_{\rm grid}$ \;
$\tilde{\tau}_{y,z} = \argmin_{\tau_{grid}} \lVert \tau_{y,z}- \tau_{\rm grid} \rVert$  $\forall (y,z)$\;
%$\tilde{\tau}_{\rm el, z} = \argmin_{\tau_{grid}} \lVert \taueln - \tau_{grid} \rVert$ \;
%$\tilde{\phi}_{\rm az, y} = \argmin_{\phi_{grid}} \lVert \phiazn - \phi_{grid} \rVert$ \;
$\tilde{\phi}_{ y, z} = \argmin_{\phi_{grid}} \lVert \phi_{y_z} - \phi_{\rm grid} \rVert$ $\forall (y,z)$\;
%Quantize  $\phiazn$, $\phieln$ to the closes value in $\phi_{\rm grid}$ \;

not-conv $\leftarrow$ True\;

\While { not-conv }{
    $G_{l, {\rm first}}$ $\leftarrow$ Eq. \eqref{eq:gl} by $\tilde{\bm\tau}$ and $\tilde{\bm\phi}$\;
    \For{y = 1:1:$N_{\rm az}$}{
    \For{z = 1:1:$N_{\rm el}$}{
        $\tilde{\tau}_{y,z} \leftarrow  \argmax_{\tilde{\tau}_{y,z} 
 \in \tau_{\rm grid} } G_{l} ( \cdots,\tilde{\tau}_{y,z},  \cdots   ) $
 }
    }

    \For{y = 1:1:$N_{\rm az}$}{
    \For{z = 1:1:$N_{\rm el}$}{
        $\tilde{\phi}_{y,z} \leftarrow  \argmax_{\tilde{\phi}_{y,z} 
 \in \phi_{\rm grid} } G_{l} ( \cdots,\tilde{\phi}_{y,z},  \cdots   ) $
 }
    }
    $G_{l, {\rm later}}$ $\leftarrow$ Eq. \eqref{eq:gl} by $\tilde{\bm\tau}$ and $\tilde{\bm\phi}$\;
    \lIf {$|G_{l, {\rm later}} -G_{l, {\rm first}}| < \zeta \times G_{l, {\rm later}} $}{not-conv=False\;}}
%Return $\tilde{\bm\tau}$, $\tilde{\bm\phi}$\;
\end{algorithm}

\subsubsection{Separated}
Similarly to the abovementioned approach, we update the parameters by searching within the defined space for phase and delay values. However, in this case, we initialize the parameters as $ \bm\tau_{\rm az}$, $\bm\phi_{\rm az}$, $\bm\tau_{\rm el}$ and $\bm\phi_{\rm el}$, according to Sec.~\ref{sec:sep_analytical}. The update process now handles $
N_{\rm az} + N_{\rm el} $ updates, compared to the prior approach's $N_{\rm az} \times N_{\rm el}$.

 %where ${\rm linspace(x,y,z)}$ leads to $z$ linearly spaced elements between $x$ and $y$. 

\subsection{Gradient Descent Algorithm}
\label{sec:gradient}
Although the greedy algorithm is guaranteed to converge to a local optimal solution, it takes a long time to do so even for the separated case.
The primary objective of the gradient descent algorithm is to quickly explore the solution space to achieve an improved beamforming gain. 
%We engage in two distinct optimization strategies: the joint and separated methodologies.

We approach the solution by using an iterative optimization technique that updates the parameters according to their gradients from the defined loss function, as in
\begin{equation}
    F (\bm\tau, \bm\phi) \triangleq \lVert G_{ l,\rm max} - G_l (\bm\tau, \bm\phi)  \rVert^2,
\end{equation}
where $G_{l \rm , max} \triangleq  10\log_{10} \left(N_{\rm az} N_{\rm el}\right)$ is the maximum beamforming gain. 
During this optimization, we schedule learning using Adam optimizer with a learning rate of $0.1$, which is determined by exploring various learning rates using a grid search. We apply this method to both solution spaces: joint and separated optimizations. 

\subsubsection{Joint}
For the joint case, the optimization parameters are $\bm\phi$ and $\bm\tau$, initialized according to \eqref{eq:phi} and \eqref{eq:tau} similar to greedy optimization. After initialization, we quantize these values with a specific precision for delay, $\tau_p$, and bit size for phase, $\beta$. Then, these parameters, $\bm\phi$, and $\bm\tau$, are updated using their gradients until meeting the convergence criteria, which is the same as Line 18 of Algorithm~\ref{alg:1}. 

\subsubsection{Separated}
In the separated case, the optimization parameters are $\bm\tau_{\rm az}$, $\bm\tau_{\rm el}$, $\bm\phi_{\rm az}$, $\bm\phi_{\rm el}$, which are calculated according to the Sec.~\ref{sec:sep_analytical} and quantized with the delay precision of $\tau_p$ and $\beta$ bits for phase. Subsequently, these parameters are modified by applying their gradients until the convergence. 

%\begin{equation}
%    G_{\rm o, max} =
%    \begin{cases}
%        N_{\rm az} N_{\rm el} & \text{o = mean} \\
%         10\log_{10} N_{\rm az} N_{\rm el} & \text{o = log-mean} \\
%    \end{cases}
%\end{equation}

 %Note that we investigate two different loss functions, \textit{mean} and \textit{log-mean}.

%The initialization of the gradient descent is the closed-form solution given in Eqs. \eqref{eq:phi} and \eqref{eq:tau}. 

\section{Simulation Results}
\label{sec:sim}

\begin{figure}[t]
\centering
\begin{subfigure}{\linewidth}
    \centering
    \includegraphics[width=0.9\textwidth]{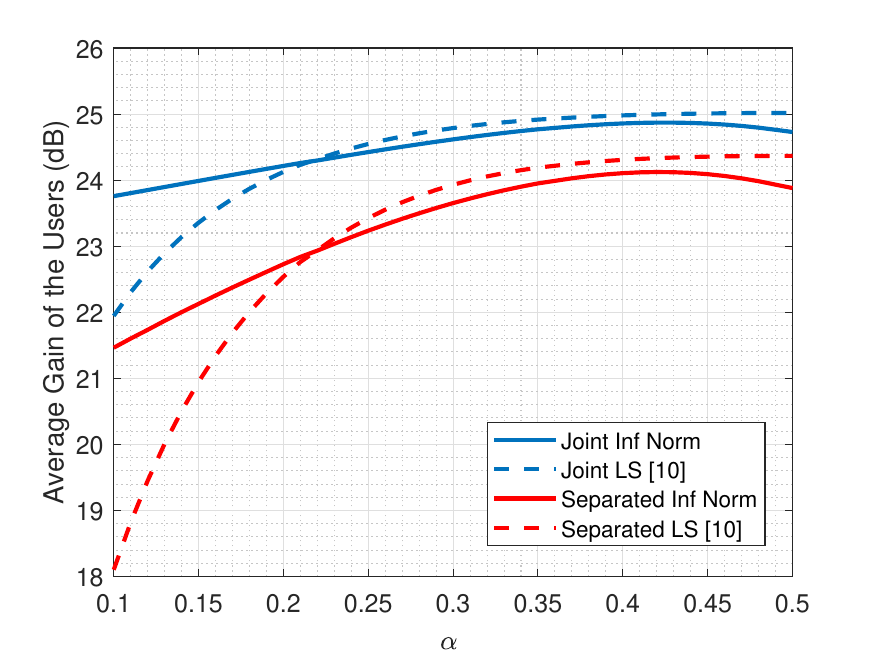}
    %\caption{Average gain of the users vs various bandwidth allocations for $N_u =2$}
    %when the angle of arrivals are $(\theta_{\rm az,1}, \theta_{\rm el,1})= (-60^\circ,90^\circ) $ and $(\theta_{\rm az,2}, \theta_{\rm el,2})= (60^\circ,120^\circ)$. }
    \caption{2-user}
    \label{fig:analytical_alpha}
\end{subfigure}
\begin{subfigure}{\linewidth}
    \centering
    \includegraphics[width=0.9\textwidth]{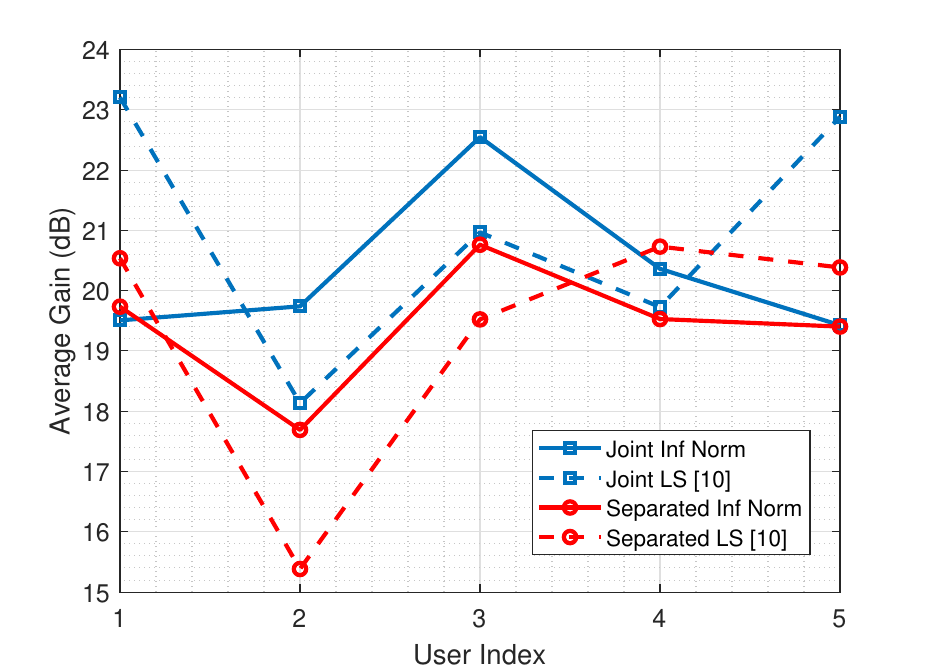}
    %\caption{Average gain vs user index for unequal bandwidth allocations} %with $\bm\alpha = [0.3, 0.2, 0.15, 0.1, 0.25 ]$ for $N_u=5$. }% for the angle of arrivals of $(\theta_{\rm az, i}, \theta_{\rm el, i}) = ((-60 + 30i)^\circ , (90 + 7.5i)^\circ )$  for $i \in[1,5]$. }
    \caption{5-user}
    \label{fig:uneq_analytical.eps}
\end{subfigure}
%\caption{Analyzing various analytical approaches with different scenarios}
\caption{Performance comparison of joint and separated analytical derivations
by using LS or infinity norm definition with a) various BW allocation scenarios and b) an unequal BW allocation for $N_u=5$ and $\bm\alpha = [0.3, 0.2, 0.15, 0.1, 0.25 ]$.}
\label{fig:set_analytical}
\end{figure}

% \begin{figure*}[t]
% \centering
% \begin{minipage}[c]{0.62\linewidth}
% \begin{subfigure}{0.49\linewidth}
%     \centering
%     \includegraphics[width=1
% \textwidth]{Figures/Upd_Figures/max_analytical_sep.eps}
%     %\caption{Separated LS }
%     \caption{Separated LS \cite{jain2023mmflexible}}
%     \label{fig:max_analytical_sep}
% \end{subfigure}
% \begin{subfigure}{0.49\linewidth}
%     \centering
%     \includegraphics[width=1\textwidth]{Figures/Upd_Figures/max_analytical_joint.eps}
%     %\caption{Joint LS}
%     \caption{Joint LS \cite{jain2023mmflexible}}
%     \label{fig:max_analytical_joint}
% \end{subfigure}
% \caption{Illustration of the maximum beamforming gain in azimuth and elevation domain for a) separated LS, and b) joint LS when $N_u=5$ with equal BW allocation.}
% \label{fig:set_max_gain}
% \end{minipage}
% \quad
% \begin{minipage}[c]{0.31\linewidth}
% \begin{subfigure}{\linewidth}
%     \centering
%     \includegraphics[width=1\textwidth]{Figures/Upd_Figures/gain_freq.eps}
%     %\caption{  $\theta_{\rm el} = 105^\circ$  }
%     \label{fig:gain_freq}
% \end{subfigure}
% \caption{ Beamforming gain of joint LS across an azimuth vs frequency domains when $\theta_{\rm el} = 105^\circ$.}
% \label{fig:set_gain}
% \end{minipage}
% \end{figure*} 

\begin{figure}[t]
\centering
\begin{subfigure}{0.9\linewidth}
    \centering
    \includegraphics[width=1
\textwidth]{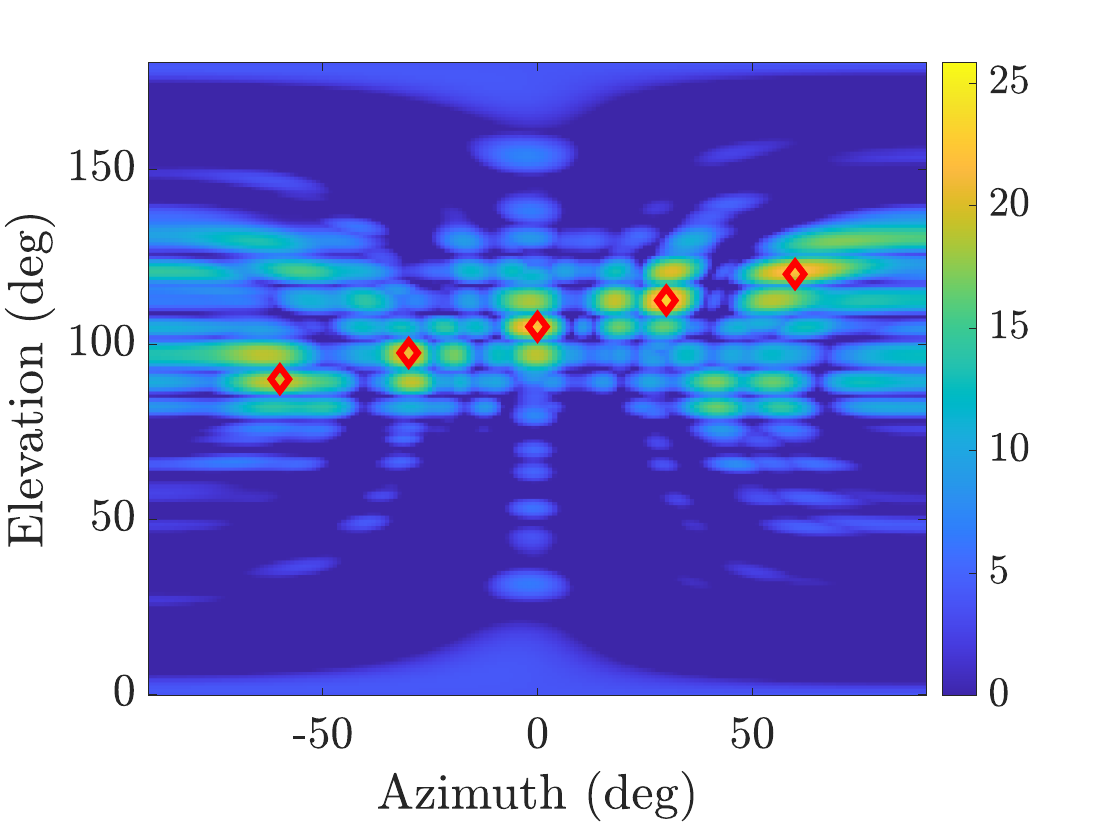}
    %\caption{Separated LS }
    \caption{Separated LS \cite{jain2023mmflexible}}
    \label{fig:max_analytical_sep}
\end{subfigure}
\begin{subfigure}{0.9\linewidth}
    \centering
    \includegraphics[width=1\textwidth]{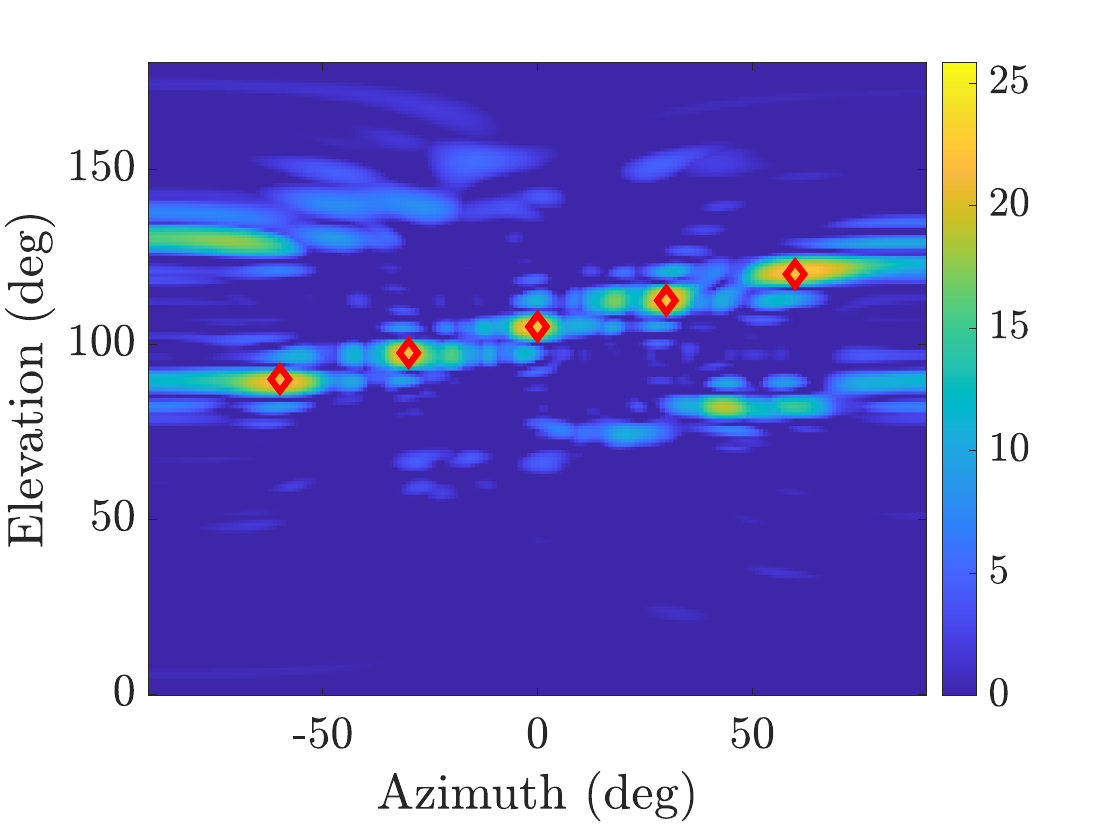}
    %\caption{Joint LS}
    \caption{Joint LS \cite{jain2023mmflexible}}
    \label{fig:max_analytical_joint}
\end{subfigure}
\caption{Illustration of the maximum beamforming gain in azimuth and elevation domain for a) separated LS, and b) joint LS when $N_u=5$ with equal BW allocation.}
\label{fig:set_max_gain}
\end{figure} 

\begin{figure}[t]
 \centering
    \includegraphics[width=0.9\linewidth]{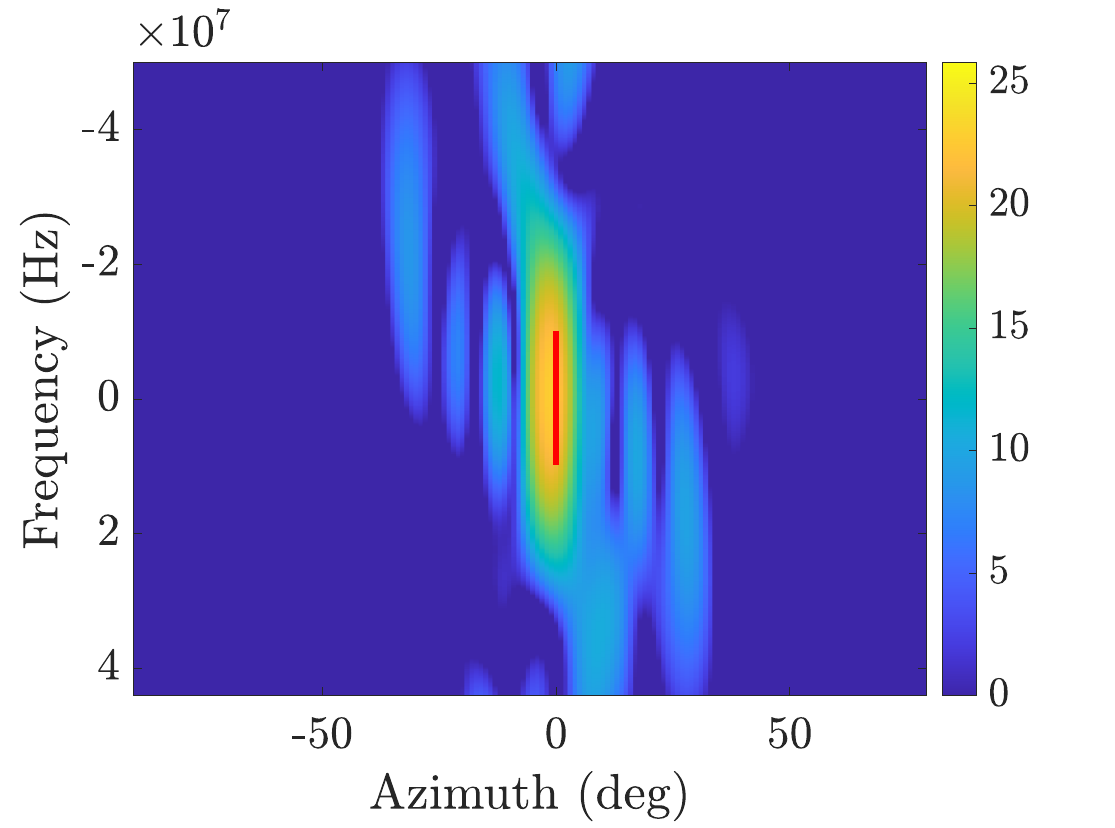}
    %\caption{  $\theta_{\rm el} = 105^\circ$  }
    \label{fig:gain_freq}
    \caption{ Beamforming gain of joint LS across an azimuth vs frequency domains when $\theta_{\rm el} = 105^\circ$. User 3 is located at $(\theta_{\rm az}, \theta_{\rm el}) = (0^\circ,105^\circ)$, and assigned the frequency band $[-9.5, 9.5)$ MHz.}
\label{fig:set_gain}
\end{figure}

\begin{figure}[t]
\centering
%\begin{minipage}[c]{0.64\linewidth}
\begin{subfigure}{0.9\linewidth}
    \centering
    \includegraphics[width=0.9\textwidth]{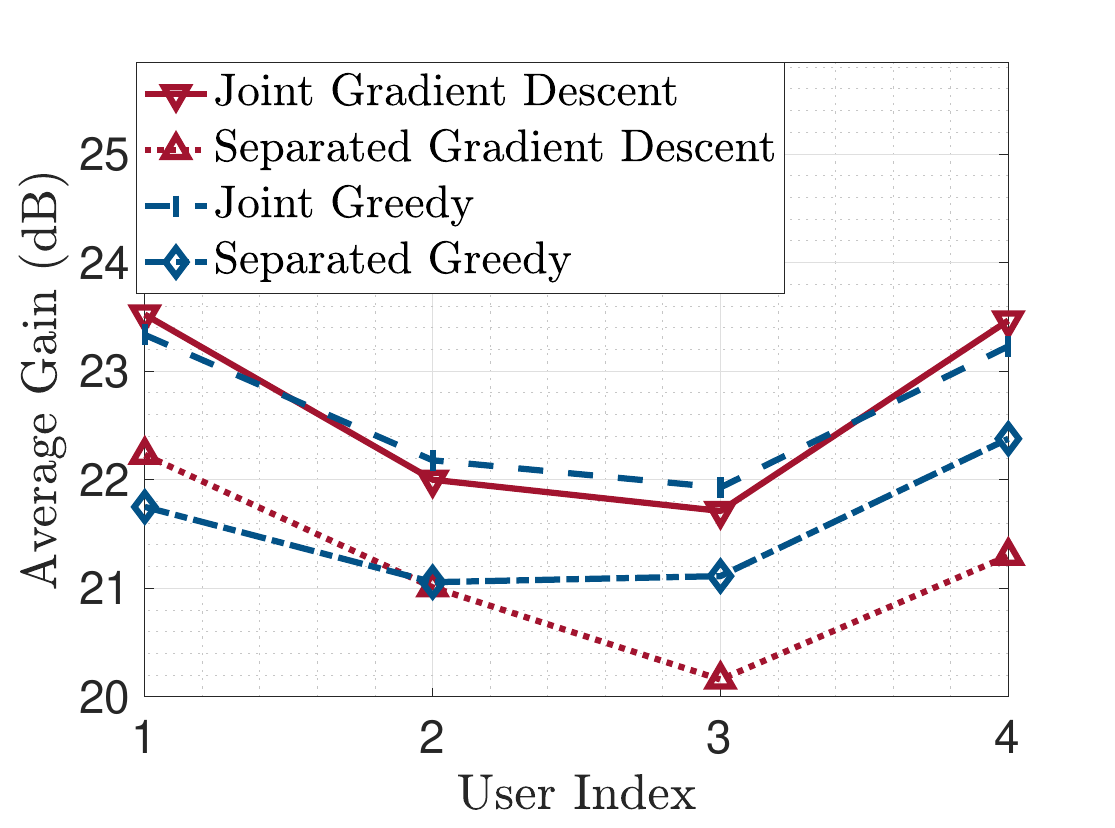}
    \caption{4-user, equal BW allocation}
    \label{fig:joint_sep_method}
\end{subfigure}
\begin{subfigure}{0.9\linewidth}
    \centering
    \includegraphics[width=0.9\textwidth]{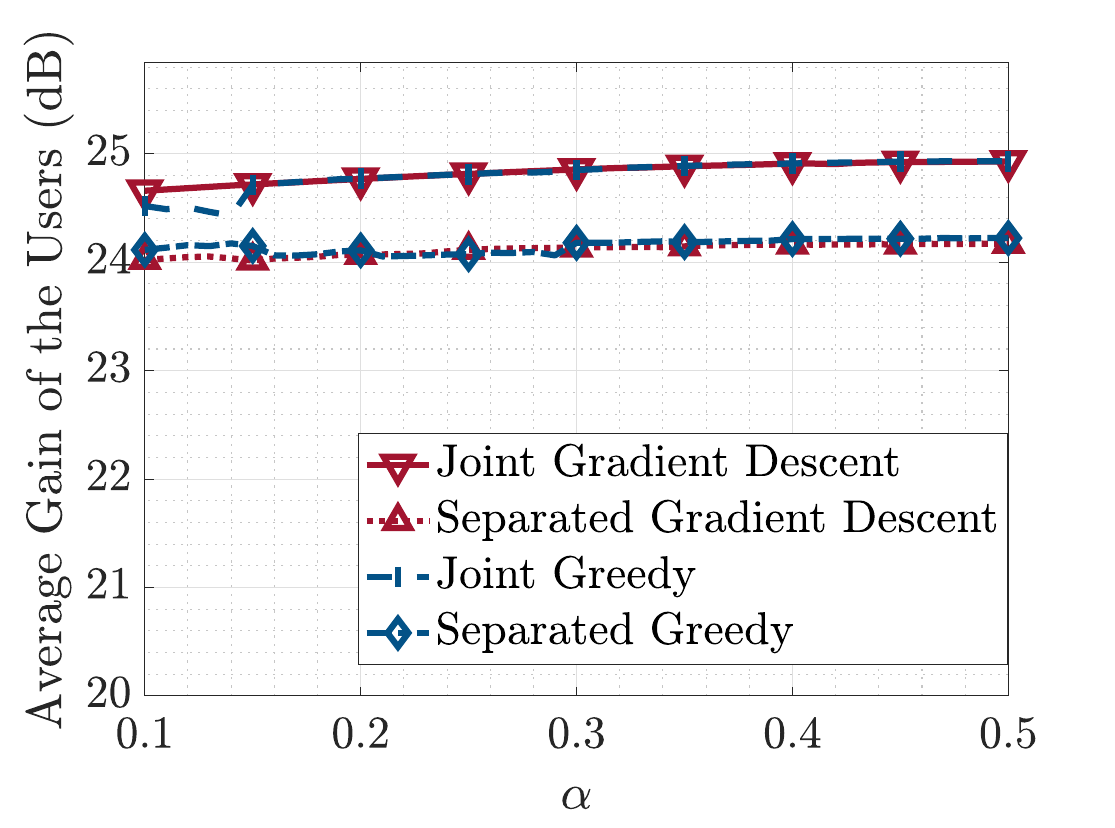}
    \caption{2-user, unequal BW allocation}
    \label{fig:joint_sep_bw.eps}
\end{subfigure}
\caption{Performance comparison of greedy and gradient descent algorithms using joint and separated optimization with a) an equal BW allocation for $N_u=4$ users, and b) various BW allocation scenarios for $N_u=2$ users. Greedy and gradient descent algorithms yield similar performance.}
\label{fig:set_methods}
%\end{minipage}
\quad
%\begin{minipage}[c]{0.33
%\linewidth}
\end{figure} 

%\begin{subfigure}{0.98\linewidth} 
\begin{figure}[t]
 \centering
    \includegraphics[width=0.9\linewidth]{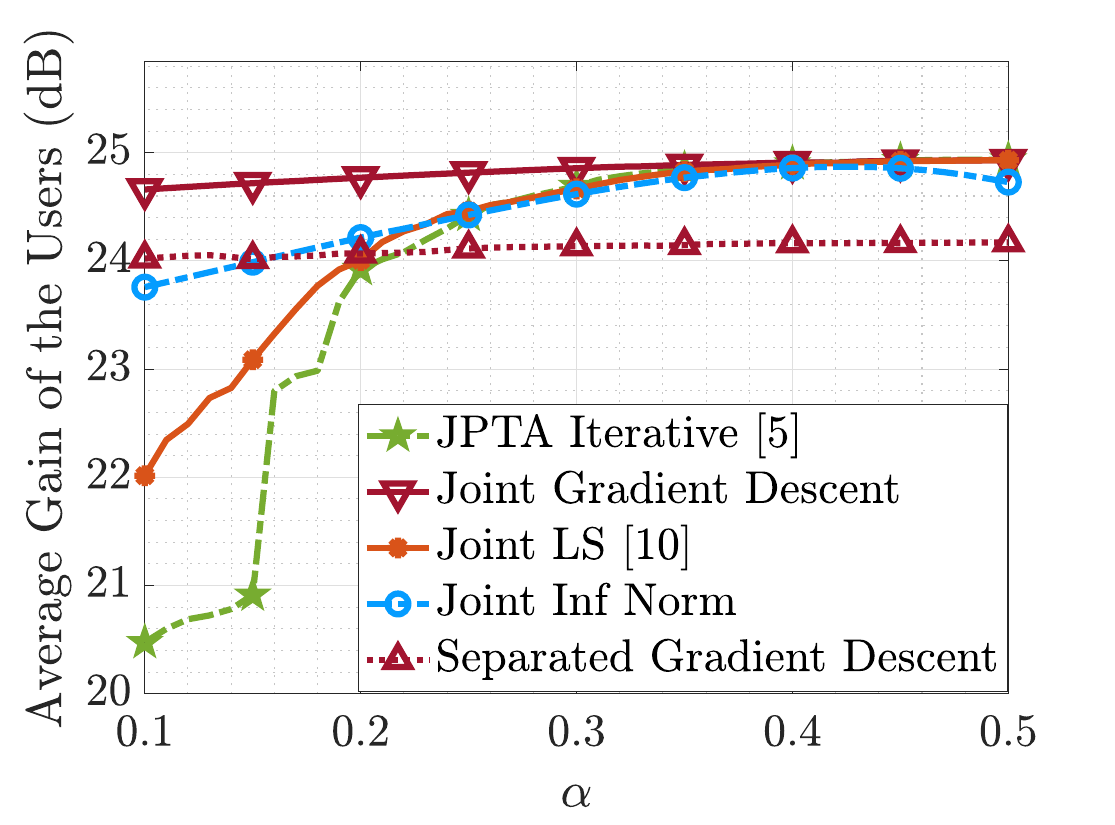}
    %\caption{Various ${\rm BW}$ allocations}
    \label{fig:comp_final.eps}
%\end{subfigure}
\caption{Comparison of our proposed solutions with the state-of-the-art over various BW allocation scenarios }
\label{fig:comp}
\end{figure}
%\end{minipage}

%\vspace*{-0.2cm}
This section provides a numerical evaluation of the methods described in Sec.~\ref{sec:method}. The simulation parameters, which are consistent with a practical 5G mmWave deployment, are given in Table~\ref{tab:tab1}. For simplicity, we assign $i$-th user's AoA as $(\theta_{{\rm az}, i}, \theta_{{\rm el}, i}) = ((-60 + i\frac{120}{N_u-1})^\circ, (90 + i\frac{30}{N_u-1} )^\circ)$ when there are $N_u$ users in a cell with $120^\circ$ horizontal coverage and $30^\circ$ vertical coverage. Note that we subject the output of the functions, phase, and delay values to quantization before illustrating their results. We present our findings of both analytical (Sec.~\ref{sec:analyt_res}) and iterative (Sec.~\ref{sec:iterative}) results. Also, we present a run-time comparison between our proposed method and state-of-the-art \cite{ratnam2022joint}.

\subsection{Results of analytical algorithms}
\label{sec:analyt_res}
In Secs.~\ref{sec:joint} and \ref{sec:sep_analytical}, we present our analytical methods to configure PS and TTD units. In \figref{fig:analytical_alpha}, we illustrate the average gains of the users, which is denoted as $G_l$ in \eqref{eq:gl}, for $N_u=2$, where User 1 gets allocated $\alpha$ of the total BW (${\rm BW}$), and User 2 gets $1-\alpha$ of the BW. Our findings reveal that joint optimization consistently outperforms separated optimization by a margin ranging from $0.6$ dB to $4$ dB, depending on the BW allocation ratio. The mean difference between joint and separated optimization, observed across various BW allocations, is approximately $1.29$ dB when the method is least squares (LS), which is an extension of \cite{jain2023mmflexible} from 2D to 3D, and $1.18$ dB when the method is infinity norm. 
Hence, it can be deduced that the greater number of systems of equations in the analytical derivation yields higher beamforming gain at the price of higher complexity. 

Moreover, the infinity-norm solution results in a higher average gain than LS for unequal BW allocation ($\alpha \leq 0.2$) in both joint and separated optimization scenarios. Although it has a higher computational complexity, the infinity-norm solution treats the two users more fairly and can achieve a larger sum data rate.
%The difference between the average gain is at most $1.81$ dB and on average $0.76$ dB over different BW allocations for the joint and at most $3.36$ dB and on average $1.4$ dB for the separated optimization.
%We can deduce the reason for this gap between LS and infinity norm methods 

In \figref{fig:uneq_analytical.eps}, we demonstrate $G_{{\rm mean}, i}$ for $i^{\rm th}$ user, defined in \eqref{gmean}, when $N_u=5$. The BW allocations of the users are $\bm\alpha = [0.3, 0.2, 0.15, 0.1, 0.25 ]$. In joint optimization, the infinity norm method yields a maximum gain difference of $3$ dB among users, whereas the LS method results in a $5$ dB difference. A similar trend is observed in separated optimization, with the infinity norm leading to a maximum difference of $3.2$ dB, while LS results in a $5.4$ dB difference. This suggests that the infinity norm method fosters greater fairness in user gain distribution, even when their assigned ratios are unequal. Furthermore, due to the more equitable distribution of average gains among users, the infinity norm also elevates the average gain of users, as illustrated in \figref{fig:analytical_alpha}.

Furthermore, we illustrate the impact of joint and separated optimization on beamforming gain in \figref{fig:set_max_gain} when the solution is obtained by LS. Here, we demonstrate the maximum gain among subcarriers for every azimuth and elevation angle when the desired user locations are marked with red diamonds. More sidelobes are observed in the case of separated optimization, as depicted in \figref{fig:max_analytical_sep}, in contrast to the joint optimization shown in \figref{fig:max_analytical_joint}. This difference can be attributed to the interdependence of the phase and delay elements, which is fully captured in the joint analytical solution. In \figref{fig:set_gain}, we display the gain across various subcarriers while maintaining a constant elevation angle, which corresponds to the User 3 in \figref{fig:max_analytical_joint} with $(\theta_{\rm az}, \theta_{\rm el}) = (0^\circ,105^\circ)$. Here, the red line is the desired maximum gain allocation from -9.5 MHz to 9.5 MHz, which is fairly satisfied by using joint optimization.

\subsection{Results of iterative algorithms}
\label{sec:iterative}

This segment presents a performance evaluation of the algorithms introduced in Secs.~\ref{sec:greedy} and~\ref{sec:gradient}. \figref{fig:set_methods} compares the greedy and gradient descent algorithms with joint and separated optimization scenarios. It's worth noting that the disparities in beamforming gains between separated and joint optimizations exhibit a consistent pattern, as discussed previously.

In \figref{fig:joint_sep_method}, we present the average gains for each user when their BW allocation is uniform. Through joint optimization, both the greedy and gradient descent algorithms achieve an average gain increase of $2.26$ dB and $2.11$ dB, respectively. A similar trend is observed when considering different BW allocations in \figref{fig:joint_sep_bw.eps}. It's worth noting that the greedy and gradient descent algorithms yield comparable performance. The primary distinction between these two methods lies in their runtime, with the gradient descent algorithm running approximately 250x faster in joint optimization cases.

In the realm of 2D frequency-dependent beam design, the algorithm proposed by \cite{ratnam2022joint} stands as the current state-of-the-art. We have extended their iterative algorithm to 3D for the sake of comparative analysis. The results of this comparison are depicted in \figref{fig:comp} as JPTA Iterative, which are across various BW allocations for a scenario with $N_u=2$. It's important to highlight that when the disparity in BW allocation is relatively small ($\alpha > 0.35$), all algorithms employing the joint optimization scheme tend to converge to the same outcome. This observation also underscores a fundamental limitation of the separated optimization approach.

Moreover, our one-shot analytical solution surpasses the iterative state-of-the-art approach proposed by \cite{ratnam2022joint} for scenarios with unequal BW allocations ($\alpha \leq 0.35$), achieving up to a $1.54$ dB gain. For $\alpha \leq 0.2$, even the gradient descent algorithm with separated optimization outperforms the other methods, except for the gradient descent algorithm using joint optimization. In conclusion, the gradient descent algorithm with joint optimization consistently performs well across various BW allocations, thus offering more reliable results. %while running faster than alternative iterative algorithms. 
When the BW is divided fairly, it can achieve a performance similar to the gradient descent algorithm using a one-shot joint analytical derivation with LS.  

\begin{figure}[t]
    \centering
    \includegraphics[width=0.85\linewidth]{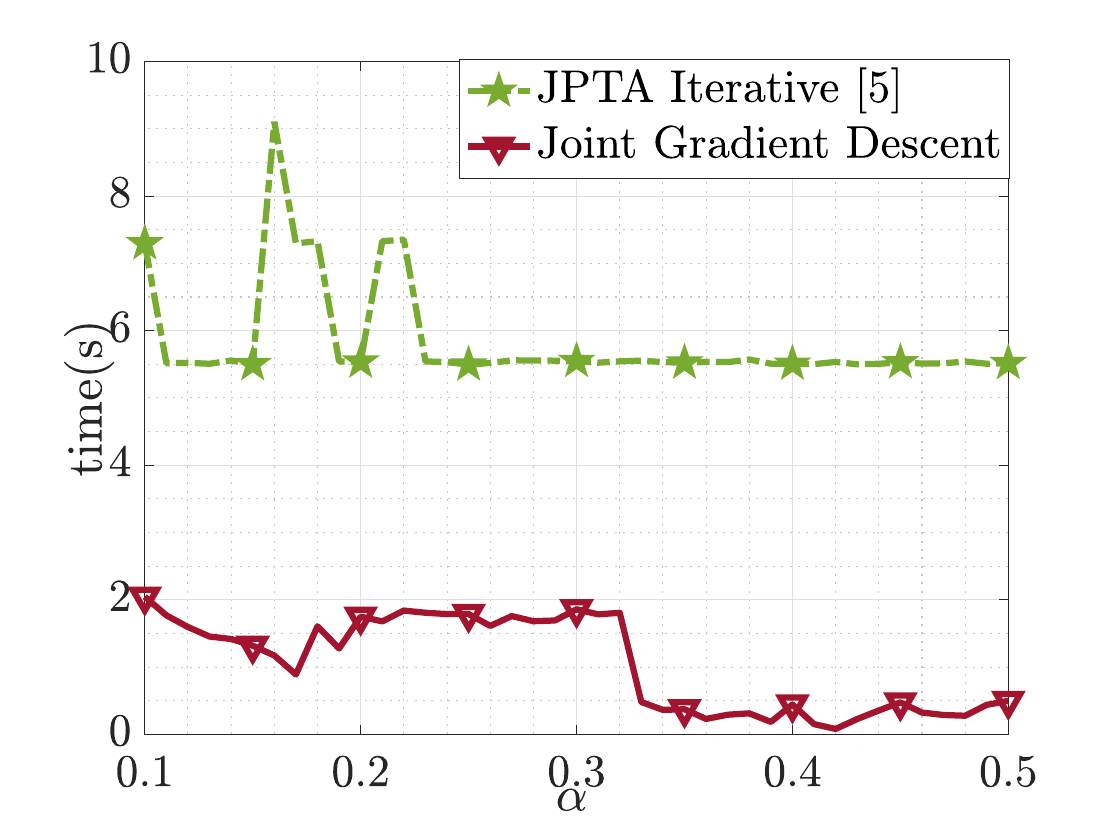}
    \caption{Time complexity of the proposed gradient descent method and the state-of-the-art iterative algorithm for different BW allocation scenarios. The codes are run on a server with Platium 8268 CPU @2.9GHz, and 2TB memory.}
    \label{fig:time_comp}
\end{figure}

\subsection{Time Complexity Analysis}
In \figref{fig:time_comp}, we demonstrate the time complexity of the iterative algorithm by \cite{ratnam2022joint} and our proposed gradient descent method with different BW allocations scenarios. It's important to note that run time of both algorithms are higher in average with unequal BW allocation ($\alpha \leq 0.35$). Since gradient descent does not improve upon the LS for equal BW allocation scenarios (see \figref{fig:comp}), the run-time is negligible. Lastly, average run-time of the gradient descent ($1.05$ s) is at least 5 times faster than the iterative algorithm ($5.8$ s).

%\Ozi{Go back to the slides to see if any conclusion is missing}

\section{Conclusion}
In this paper, we have investigated the 3D frequency-dependent beam design with joint phase-time arrays, aiming to serve multiple users in a single shot and thereby reduce beam search and scheduling latency.
We have presented analytical derivations to maximize the beamforming gain and then bolstered our findings with two iterative algorithms: greedy and gradient descent. 

By simulating the 3D beamforming gain, we have compared the performance of our proposed algorithms with the state-of-the-art, originally designed for 2D and extended to 3D by us. Our findings indicate that our proposed gradient descent algorithm surpasses the state-of-the-art, delivering more reliable results across various scenarios. Furthermore, our joint analytical derivation matches the performance of the gradient descent algorithm when the bandwidth allocation is relatively uniform. Lastly, the simulation results clearly show that the separated optimization, usually adopted for the UPA 3D beam design, does not work well for the JPTA use case, since it leads to lower gain and more sidelobes.

Besides the beam design, a big challenge for the actual implementation of 3D JPTA is the aperture size and cost of delay elements. The number of delay elements increases linearly with the number of antennas in both azimuth and elevation directions in the considered structure. A future direction is to assign fewer delay elements in the elevation direction by connecting multiple antennas to a single delay element. 
%For example, reduce the number of delay elements from $N_{\rm az} N_{\rm el}$ to $N_{\rm az}$ by connecting all the antennas in the same column to a delay element. 
Because the users are more distributed in the azimuth direction in the outdoor cellular network, this simplified design can still achieve most of the benefits of 3D beamforming. 

\bibliographystyle{IEEEtran}
\bibliography{ref}
\end{document}